\documentstyle[12pt,aasms4]{article}

\begin{document}
\title {One-Armed Spiral Waves in Galaxy
Simulations with Counter-Rotating Stars} \smallskip \normalsize

{\large One-Armed Spiral Waves in Galaxy
Simulations with Counter-Rotating Stars}\\

Neil F.\ Comins, Department of
Physics and Astronomy, Bennett Hall, University of Maine,
Orono, ME 04469, galaxy@maine.maine.edu\\

Richard V.\ E.\ Lovelace, Department of
Astronomy, Cornell University, Ithaca, NY, 14853, rvl1@cornell.edu\\

and\\

Thomas Zeltwanger \& Peter Shorey, Department of
Physics and Astronomy, Bennett Hall, University of Maine, Orono, ME 04469\\

\begin{abstract}
Motivated by observations of disk
galaxies with counter-rotating stars,
we have run two-dimensional,
collisionless $N-$body simulations of disk
galaxies with significant counter-rotating
components. For all our simulations the
initial value of Toomre's stability
parameter was $Q = 1.1$. The percentage of counter-rotating
particles ranges from $25\%$ to $50\%$.
A stationary one-arm spiral wave is
observed to form in each run, persisting
from a few to five rotation periods,
measured at the half-mass
radius. In one run, the spiral
wave was initially a leading arm which
subsequently
transformed into a trailing arm.
We also observed a change in spiral
direction in the run
initially containing equal
numbers of particles orbiting in
both directions. The results of our simulations
support an interpretation of the
one armed waves
as due to the two stream instability.
\end{abstract}
\keywords {gravitation---instabilities---galaxies:
evolution---kinematics \& dynamics---stars:kinematics}

\section{Introduction}
Several disk galaxies have been
observed to contain counterrotating
stars, including NGC 3593
\markcite{bertola1996}
(Bertola et al. 1996); NGC 4138
\markcite{Jore1996} (Jore, et al. 1996),
\markcite{Thakar1996} (Thakar, et al., 1996);
NGC 4550 \markcite{Rubin1992}
(Rubin, et al. 1992)
\markcite{Rix1992}, (Rix et al. 1992);
NGC 7217 \markcite{Merrifield1994}
(Merrifield \& Kuijen 1994);
and NGC 7331 \markcite{Prada1996}
(Prada, et al. 1996).
The observed mass fraction
in counterrotating stars is
remarkably high, ranging
from $\approx 20\%$ to
$\approx 50\%$. Lovelace, Jore, and
Haynes (1997)
\markcite{lovelace1997}(hereafter LJH)
developed a theory of
the two-stream instability in
flat counter-rotating
galaxies. The basic
instability is similar to
that found in counterstreaming
plasmas
\markcite{Krall1973}
(e.g., Krall \& Trivelpiece 1973).
LJH made several predictions,
most notably that the presence
of stars orbiting in both
directions around the disk
should create a strong instability of
the
$m=1$ (one-arm) spiral waves;
that the $m=1$ wave with the strongest
amplification is usually
the leading spiral arm
with respect to the dominant
component;
and that the
spiral wave is stationary
when there are equal co- and counterrotating components.
The two-stream instablity may have
been observed in earlier computer
simulations which found $m=1$ spiral
waves in counterrotating disks
(Sellwood \& Merritt 1994;
Sellwood \& Valluri 1997; Howard et al. 1997);
however the characterization the
instability was very limited and
tightly wrapped spiral waves were not
observed. It is of interest
that $m=1$ perturbations in spiral
disks are rather common
(Rix \& Zaritsky 1995; Zaritsky \& Rix 1997).
In some cases these may arise from
counterrotating material.

We present here the results of
three runs of our
GALAXY code \markcite{SchroederComins1989}
(Schroeder \& Comins 1989;
\markcite{Schroeder1989}
Schroeder 1989; \markcite{Shorey1996}
Shorey 1996), each with different
fractions of the
stars initially in counterrotating orbits.
The runs had $25\%$, $37.5\%$,
and $50\%$ counterrotating
particles. In the notation
of LJH, this corresponds to
runs with $\xi_* = 0.25$,
0.375, and 0.50, respectively.
Each run had a total of
100,000 collisionless,
equal-mass particles embedded
in a halo containing $75\%$ of
the total mass of the system.
This halo mass is used to help
suppress the ubiquitous $m=2$
``bar-mode'' instability.
This particular halo mass
fraction is chosen because
it restricts the bar to the
inner one-quarter of the
disk for the run with all
the particles moving in
the same direction, while, along
with the effect of the
counterrotating particles,
helping to completely
suppress the bar
in the
$\xi_* = 0.50$ run \markcite{Kalnajs1977}
(Kalnajs 1977). A halo mass fraction much
larger than this severely limits
the instability we are studying
\markcite{Comins1997} (Comins et al. 1997).

The simulation is done on a
Cartesian grid with
$256 \times 256$ cells.
The radial mass distributions
of both the particles and the
halo are that described in
\markcite{Sellwood1984}
Sellwood \& Carlberg (1984)
and \markcite{Carlberg1985}
Carlberg \& Freedman (1985)
for simulating the rotation
curve of an Sc galaxy. The
initial value of Toomre's (1964)
stability parameter is $Q = 1.1$
over the entire disk.  This Q increases due to heating
throughout the run,
bringing the disk to a Q consistent with
the values in real galaxies.
Toomre's critical radial wavenumber
$k_{crit}=\kappa^2/(2\pi G \Sigma)$
satisfies the condition assumed by LJH for
tightly wrapped spiral waves, $k_{crit} r \gg 1$, over
all but the very center of the disk ($k_{crit}r$ varies
from $7$ at $0.1r_{max}$ to $15$ at $r_{max}$).
Here,
$\kappa$ is the epicyclic frequency
and $\Sigma$ is the total surface mass density.
For our Galaxy, $k_{crit}r \approx 2\pi$ at
the radius of the Sun (Binney \& Tremaine 1987).
The LJH theory
predicts an e-folding time of about $1/\pi$ of
a rotation period for $\xi_*=0.5$ for a wave
with radial wavenumber
$(k_r) \approx 1.85 k_{crit}$ and $Q=1$.
The e-folding time increases and
the wavenumber decreases (to $(k_r)\approx
0.9k_{crit}$) as $Q$ increases.
There is no growth for $Q > 1.8$.
Each
run lasted 8 rotation periods,
as measured at the half-mass
radius of the disk. In $\S\S 2 - 4$
we consider each run separately.
In \S 5 we compare and contrast
them and present our conclusions.

\section{$\xi_* = 0.50$ Case}
This run began with very
little large$-m$ spiral development,
compared to what occurs in the
other cases we consider or when all the
stars are orbiting in the same
direction. Indeed, the disk
remains featureless for
three-quarters of a rotation
period. Theoretically, the
$\xi_* = 0.50$ case
(equal numbers of particles
traveling in both directions)
has no preferred direction of
motion. However, the symmetry
is broken by the Monte Carlo
particle position and velocity loads.
Initially there are $0.03\%$
more particles moving clockwise
in this run.

The initial, stationary spiral in this
run points counterclockwise
(Figure 1).
Stationary here, and throughout
this paper, means that the spiral structure rigidly
rotates less than $10^o$ per rotation period.
\begin{figure}[htb]
\begin{center}
\leavevmode
\epsfxsize=8cm
\epsffile{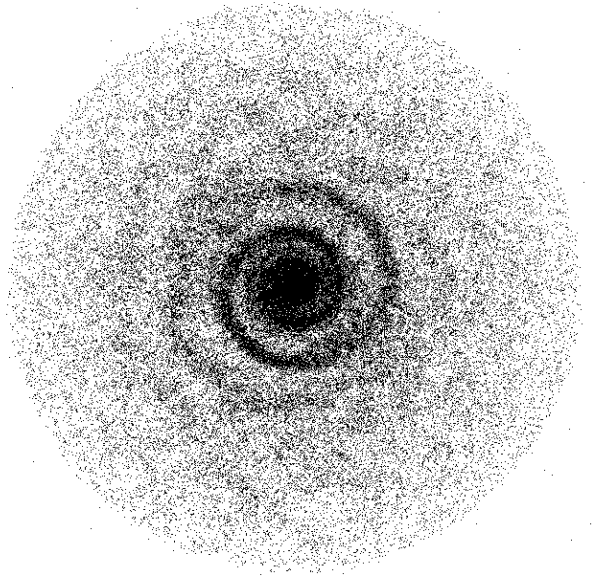}
\caption{Leading, one-arm
spiral concentrated in the inner
regions of the $\xi_* = 0.50$ run
at time
$t=1.75$ rotation periods, where
the rotation period is measured
at the half mass radius.
This spiral developed after
only $3/4$ of a rotation
period and persisted for
4 rotation periods.
For
all our runs, the radial mass
distribution models an Sc galaxy;
is $10^5$; 75\% of the
gravitating mass is in an
inert halo with the same
radial mass distribution as
the stars; the Cartesian grid
is composed of $256 \times 256$ cells;
and $Q_{0} = 1.1$.}
\end{center}
\end{figure}
The spiral shown in Figure
1 became noticeable after $3/4$ of a rotation
period of the disk and it
persisted for 4 rotation
periods.
During this time it was
stationary and
its linear growth had ceased,
giving saturation of the mode amplitude.
As defined in LJH, the initial growth rate
$\omega_i \approx
0.14 \Omega$ during the first rotation period
is well below the maximum growth rate
predicted by LJH of $0.5 \Omega$.
The
spiral arm then changed direction,
taking one rotation
period (between the fifth
and sixth rotation periods) to make the
transition.
This reversed spiral was also
stationary and it persisted
for about two rotation periods,
by which time the disk also
showed a series of spiral
arcs. Figure 2 shows the
reversed spiral.
\begin{figure}[htb]
\begin{center}
\leavevmode
\epsfxsize=8cm
\epsffile{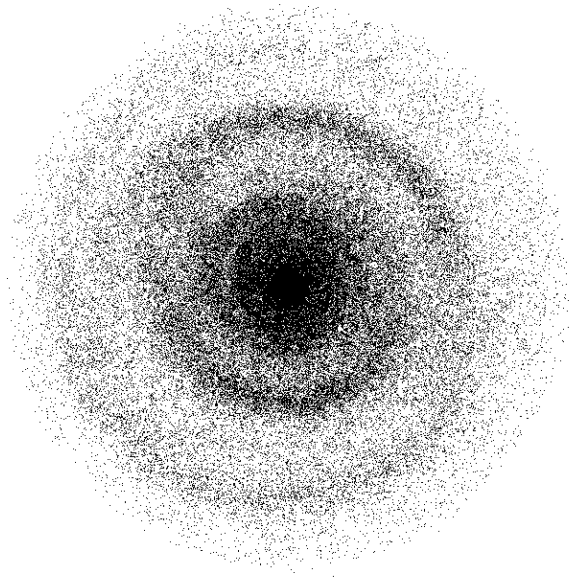}
\caption{Trailing, one-arm
spiral late in the $\xi_* =
0.50$ run at time $t=6.5$
rotation periods.
Note that unlike
the earlier, leading-arm spiral
shown in Figure 1, this spiral
covers almost the entire disk.}
\end{center}
\end{figure}
The $m=1$ Fourier
amplitude begins to decrease
as soon as $Q$ exceeded 1.8,
as predicted by LJH.
Figure 3 shows this effect.
\begin{figure}[htb]
\begin{center}
\leavevmode
\epsfxsize=8cm
\epsffile{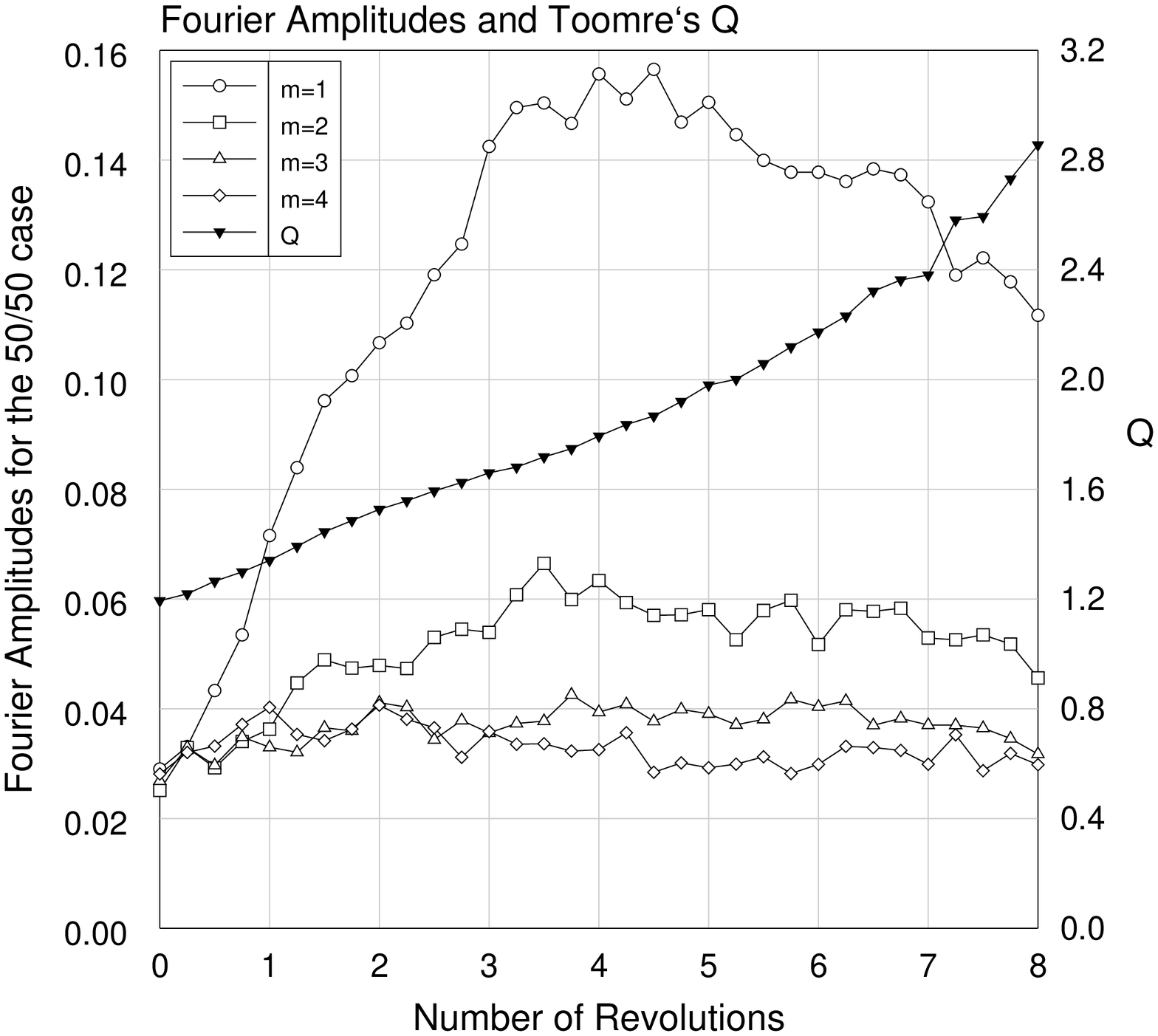}
\caption{Azimuthal Fourier
amplitudes and Toomre's $Q$
for the $\xi_* = 0.50$ run.
These results are averaged over
the entire disk. Note the
decrease in the $m=1$ Fourier
amplitude after $Q$ increases
beyond $1.8$.
The Fourier
amplitudes are $A_m(t) =
\sum_{j=0}^{95} \sum_{k=1}^{32}
\Sigma_k(j,t) e^{2 \pi i k m/32}$
where $j$ indicates the annulus and
$\Sigma_k(j,t)$ is the surface mass
density in annulus $j$ at angle $2 \pi k/32$.}
\end{center}
\end{figure}

\section{$\xi_* = 0.375$ Case}
This run begins with the development of
weak multiarm spirals.
Within
one rotation period these are
replaced by a stationary,
leading, one-arm spiral
that persists for about
two rotation periods.
This spiral is similar in
appearance to the one in Figure 1,
rotated by roughly $180$ degrees.
This spiral then transforms
into a single trailing-arm
spiral. The change occurs
in a fraction of a rotation
period. This spiral then
splits into a combination
of leading and trail-arm
segments. Figure 4 shows
the mix of leading and
trailing arcs that constitute
the stationary structure for
five rotation periods late
in the run.
\begin{figure}[htb]
\begin{center}
\leavevmode
\epsfxsize=7cm
\epsffile{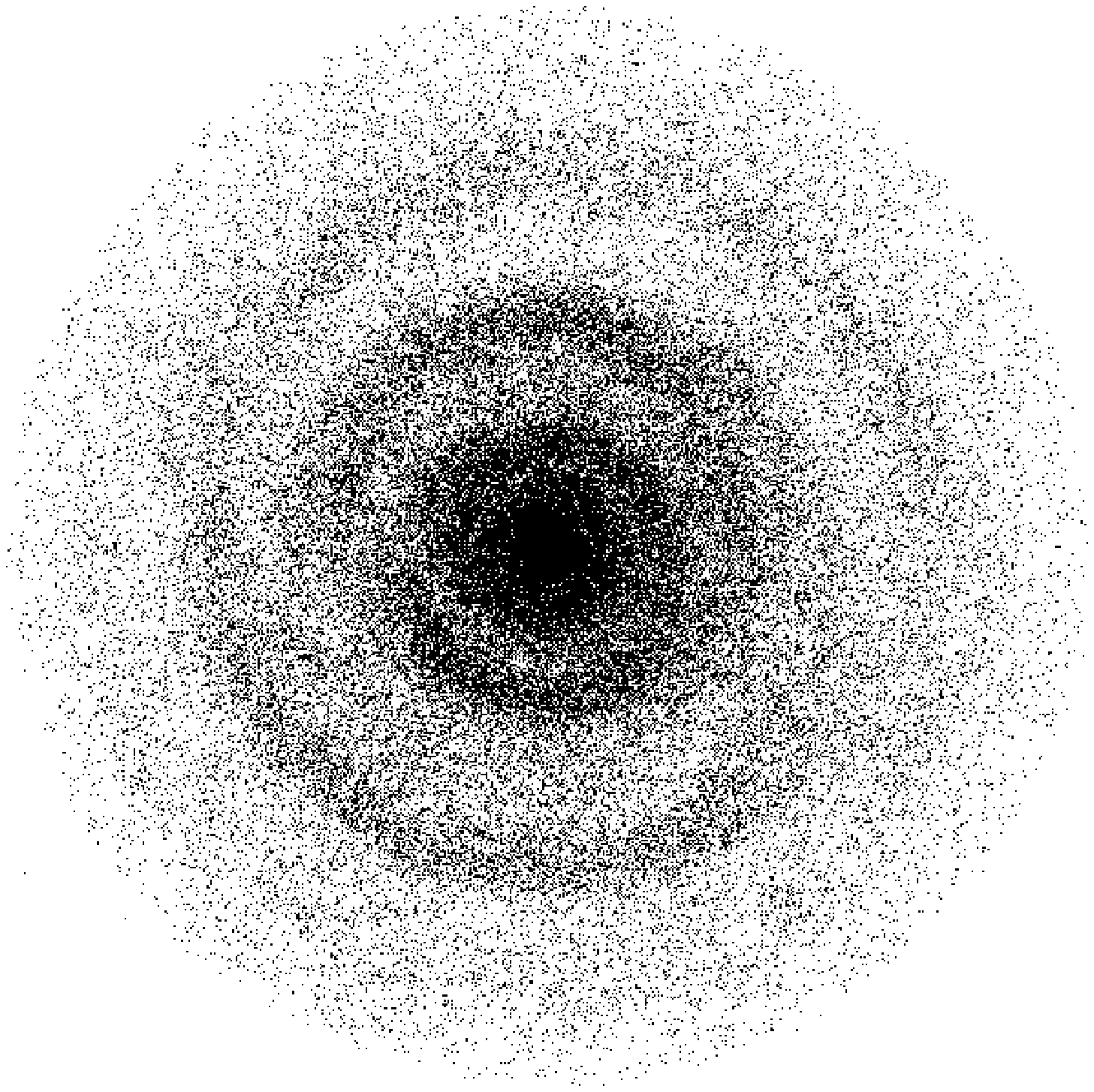}
\caption{Arcs late in the $\xi_* =
0.375$ run, at time $t=5$ rotation
periods. These developed after
the leading-arm spiral transformed
into a trailing one-arm spiral during
less than $1/4$ of a
rotation period. The trailing
spiral fragmented into the arcs
depicted here, which persisted
for 5 rotation periods. }
\end{center}
\end{figure}
Once again, the
$m=1$ Fourier amplitude begins
to decay when $Q$ increases above 1.8.

\section{$\xi_* = 0.25$ Case}
This run begins with the formation
of trailing multi-arm spirals
within the first one quarter
of a rotation period. These
arms are numerous, relatively
weak (Fourier amplitudes less
than 0.003), and they are
similar in appearance to
those seen at the beginning
of a typical run in which all
particles are orbiting in the
same direction ($\xi_* = 0$).
The initial arms persist for over
three rotation periods. During
this time, an $m=1$ trailing-arm
spiral forms, eventually
dominating the disk with a
maximum Fourier amplitude of 0.10.
The weaker multiple-arm structure
vanishes. The one-arm spiral
is stationary and it
remains for about 4 rotation periods.
It is less tightly wound than
the spiral in Figure 1, but
more tightly wound than the
spiral in Figure 2.
The $m=1$ mode begins decreasing
in Fourier amplitude as $Q$ increases
above $1.8$, as predicted by
LJH. As this $m=1$ spiral fades,
it is replaced with a small bar,
which persists for the
remainder of the run.

\section{Discussion}
Our simulations support many
of the features of the ``two-stream''
instability in galactic disks
with counterrotating stars,
as presented in LJH.
Consistent with the
theory we see:
one-armed spirals; decrease
in the amplitude of this
spiral as the Toomre $Q$ of
the system exceeds $1.8$; strengthening
of the spiral wave amplitude
with $\xi_*$ (for $\xi_* \leq 0.5$);
and leading-arm spirals.
The strength of the one-arm
instability is predicted by
LJH to be strongest in
the $\xi_* = 0.50$ case,
consistent with our simulation.
Indeed, the Fourier
amplitude $A_1$ of the arm there
is 7 times greater than in
the $\xi_* = 0.375$ case
and 16 times greater
than in the $\xi_* = 0.25$ case.

We do note some discrepencies
between our results and
the existing two-stream
theory. First, the one-armed
spiral in the $\xi_* = 0.25$
run developed and remained
as a trailing-arm spiral.
We attribute this to the
particularly strong trailing,
multiarm spirals that
developed at the beginning
of this run. We believe
that the organized motion
of the particles in the
trailing arms damped the
leading arm instability,
and transferred energy
to the more slowly amplifying,
trailing, one-arm spiral.
The influence of strong initial
trailing-arm spiral growth
and the resulting coupling
between modes with different
numbers of arms is not
addressed in the two-stream
instability theory as
presented in LJH. Therefore,
this result probably does
not contradict the theory.

Sellwood and Merritt (1994)
earlier found unstable $m=1$ modes,
but their work differs from the
present paper by having $k_r r$ not large
compared with unity and having the
unstable modes localized in the inner
part of the disk where $Q(r)$ was
smallest and where the rotation curve
was rising.   Comparisons with Sellwood
and Valluri (1997) and Howard et al. (1997)
is not possible because they do not
give $k_{crit}(r)$ .

Finally, we observed dominant
spirals changing direction, a
phenomenon that is not
predicted by the present
version of the two-stream
instability.
Figure 5 shows that with pattern
speed $\Omega_p = 0$,
the spiral encounters no resonances,
making it possible for the wave to
propagate through the center
of the disk and thereby
reverse direction.

For the $\xi_* = 0.375$ case,
the leading arm spiral occurred
only in the inner quarter of
the disk's radius. The subsequent
trailing arm extended over three
quarters of the disk.
This transformation appears to be
consistent with the behaviour of
a swing amplfier.  We see
precisely the same qualitative behaviour
as depicted in Toomre (1981 Figure 8), where the
leading spiral is only in the inner region
of the disk.  It then undergoes a transformation
to a trailing arm spiral extending over a much
larger radial extent.  We are pursuing this issue of
whether we are seeing swing amplification.

The strength and long lifetimes
of the one-arm spiral waves,
as shown by the runs presented
here, suggests that there may be
one-armed spiral features in galaxies
with counterrotating components.

We are continuing to pursue this
intriguing phenomenon in a
variety of ways, including
simulations with more
particles (which will slow the rate
of increase of $Q$), simulations using
a gravitating gas component,
simulations using other
initial values of $Q$, and simulations
using other initial radial mass
distributions. In particular,
we have found the one-arm spiral waves
to persist
for over ten rotation periods
in a $Q_o = 1.1$,
$\xi_* = 0.5$ Kuz'min disk.
We are also developing a three dimensional
code in which to explore mergers of
counterrotating disks.

\section{Acknowledgements}
N.F.C. wishes to
thank Sun Microsystems, Inc.,
for a grant of computers on
which this work was done, and
Bruce Elmegreen for a helpful discussion.
The work of R.V.E.L. was
supported in part by NSF
grant AST 93-20068.

\end{document}